\documentclass[final,5p,times,twocolumn]{elsarticle}

\usepackage{amssymb,amsmath,amsthm}
\usepackage{mathrsfs}
\usepackage{hyperref}
\usepackage{color}

\begin{document}

\begin{frontmatter}

\title{Modified gravity as a \emph{diagravitational} medium}

\author{Jos\'e A.~R.~Cembranos}
\ead{cembra@ucm.es}
\author{Mario Coma D\'iaz}
\ead{marioctk@gmail.com}
\author{Prado Mart\'in-Moruno}
\ead{pradomm@ucm.es}
\address{Departamento de F\'isica Te\'orica and UPARCOS, \\Universidad Complutense de Madrid, \\E-28040 Madrid, Spain}

\begin{abstract}
In this letter we reflect on the propagation of gravitational waves in alternative theories of gravity, which are typically formulated using extra gravitational degrees of freedom in comparison to General Relativity. We propose to understand that additional structure as forming a \emph{diagravitational} medium for gravitational waves characterized by a refractive index. Furthermore, we shall argue that the most general diagravitational medium has associated an anisotropic dispersion relation. 
In some situations a refractive index tensor, which takes into account both the deflection of gravitational waves due to the curvature of a non-flat spacetime and the modifications of the general relativistic predictions, can be defined. The most general media, however, entail the consideration of at least two independent tensors.
\end{abstract}


\end{frontmatter}



\section{Introduction}\label{Introduction}

Nobody in our community is unaware of the first direct measurement of Gravitational Waves (GWs) announced already two years ago \cite{Abbott:2016blz}.
The confirmation of this general relativistic prediction reinforced the glory of Einstein's theory and our confidence in its power to describe the gravitational phenomena.
Even so, it did not prevent researchers from continuing to investigate on alternative theories of gravity (ATGs), as they may describe the early inflationary period and provide an option to the dark energy paradigm.
Moreover, those theories, which typically have more degrees of freedom than General Relativity (GR), also predict gravitational radiation.
This radiation, however, has more than two polarizations and propagate in a different way \cite{Will:2014kxa}.
The tensorial modes correspond to what is properly called the GW. They can have a propagation speed different from the speed of light and from that of the vector or scalar modes.
Gravitational radiation offers, therefore, new routes of proving modifications of the predictions of GR or finding additional support for it.
In fact, the recent measurement of GWs and their electromagnetic counterpart \cite{TheLIGOScientific:2017qsa} has evidenced that they propagate at the speed of light in the nearby universe; hence it has allowed to rule out a large number of ATGs as dark energy mimickers  \cite{Lombriser:2015sxa,Lombriser:2016yzn,Ezquiaga:2017ekz,Creminelli:2017sry,Sakstein:2017xjx,Baker:2017hug,Akrami:2018yjz}.

In this context, it is an interesting exercise to reflect on the fundamentals of GWs in ATGs to shed some light on unexplored phenomena in these theories. The propagation of GWs in cosmological spacetimes, as predicted by different promising ATGs, has been thoroughly investigated, see e.~g.~\cite{deRham:2014zqa,Saltas:2014dha,Caprini:2018mtu,Akrami:2018yjz}. From those studies and basic knowledge on electrodynamics, it is easy to conclude that the modifications of the general relativistic predictions introduced by ATGs affect the propagation of GWs as if they were moving through a medium, as compared with the motion through vacuum where those modifications are negligible. Therefore, the additional ``structure" introduced by ATGs, which alter the dialog between matter and curvature, affects the propagation of GWs producing what we will call a \emph{diagravitational medium}, in analogy with a dielectric medium in electrodynamics.

On the other hand, it is worth mentioning that the deflection of light in curved geometries can be investigated using an effective refractive index that is a $3\times3$ tensor defined in terms of the metric \cite{Boonserm:2004wp}. In a general relativistic context, a scalar effective refractive index has already been defined to describe the deflection of GWs in some studies \cite{Szekeres,Peters,Giovannini:2015kfa}.
In this letter we will fully develop the analogy between the propagation of light through a dielectric medium and the propagation of GWs predicted by ATGs, as if they were moving through a diagravitational medium. We will consider that in the context of ATGs and general backgrounds, the effective refractive index should be a $3\times3$ tensor with entries depending on the background metric and on the new gravitational terms introduced by the ATG. Developments on the dispersion relation for electromagnetic waves in anisotropic media \cite{Itin:2009fj} suggest us that it is more appropriate to consider two independent tensors characterizing the medium instead of one refractive index.
However, we will first work in detail the simplest scenario, summarizing previous advances on ATGs, to acquire intuition about the problem. As it could be expected, in Minkowski space the effective refractive index of GWs can be characterized by a scalar quantity.
It should be noted that a first approach to include modified gravity effects in the scalar refractive index of highly symmetric backgrounds has been recently considered \cite{Giovannini:2018zbf,Giovannini:2018nkt} during the final stage of development of this work (focusing on cosmological scenarios in that case) .

\section{Propagation through isotropic diagravitational media}
Let us first consider the propagation of GWs in the simplest scenario. A large number of ATGs predict tensor perturbations that propagate in a Minkowski background according to \cite{Saltas:2014dha}\footnote{Equation (\ref{eqS}) is similar to that presented in reference \cite{Saltas:2014dha} for cosmological spacetimes, although our definition of $\nu$ is adapted to Minkowski, we have not used a Fourier decomposition for the wave, and we have not (yet) considered bigravity. Note that we neglect the backreaction of the new degree of freedom into the background geometry and  the effect of the potential mixing of their perturbations with the GWs. Moreover, we assume that the unperturbed new degree of freedom respect homogeneity and isotropy.}
\begin{equation}\label{eqS}
\ddot h_{ij}(t, \vec x)+\nu\, \dot h_{ij}(t, \vec x)-c_T^2\vec\nabla^2h_{ij}(t, \vec x)+m_{\rm g}^2\, h_{ij}(t, \vec x)=0,
\end{equation}
where $h_{ij}(t, \vec x)$ is the transverse and traceless part of the metric perturbations and we use natural units throughout this work. The term $\nu$ takes into account the potential run rate of the effective Planck mass, $c_T$ is the speed of GWs, and $m_{\rm g}$ is the graviton mass ($\nu=0$, $c_T=1$, and $m_{\rm g}=0$ in GR). Although $\nu$ and $c_T$ typically depend on the additional gravitational degrees of freedom, we assume that they are slowly varying functions along the intervals of interest and, therefore, we take them as having a constant value. 

The solution of equation (\ref{eqS}) can be expressed in terms of the frequency $\omega$ and wave number vector $\vec k$ of the wave.
We assume that
\begin{equation}\label{eqn}
k=n\,\omega,
\end{equation}
with $k=\sqrt{\vec{k}^2}$, which is just the standard definition of refractive index used in electrodynamics.
Taking a plane-wave solution of equation (\ref{eqS}), we obtain
\begin{equation}\label{nt}
n^2=\frac{1}{c^2_{T}}\left(1+i\frac{\nu}{\omega}-\frac{m_{\rm g}^2}{\omega^2}\right).
\end{equation}
This equation is exact in Minkowski space. It can also be applied to propagation through small distances of any curved geometry.
In particular it is valid during the early epochs of our Universe for which relevant constraints on the speed of GW propagation have not yet being obtained.

Now we focus on situations where the ATG implies only slight departures from the general relativistic predictions, that is $n\simeq1$. This will certainly be satisfied in the recent Universe \cite{Ezquiaga:2017ekz,Baker:2017hug,Creminelli:2017sry}.
Furthermore, we consider that $n$ can be generalized by taking into account the possibility of having a modified dispersion relation with the extra term $Ak^\alpha$ \cite{Mirshekari:2011yq},
where $A$ is a dimensionful constant. Therefore, the effective refractive index is given by
\begin{equation}\label{neff}
n=1+i\frac{\nu }{2\,\omega}+(1-c_{T})-\frac{m_{\rm g}^2}{2\,\omega^2}-\frac{A}{2}\omega^{\alpha-2},
\end{equation}
up to first order in $\nu/\omega$, $1-c_T$, $m_g^2/\omega^2$, and $A\,\omega^{\alpha-2}$. So, GWs in ATGs propagate as if they were in a medium formed by the extra gravitational degrees of freedom, whereas GWs in GR are analogous to electromagnetic waves propagating in vacuum. Using this analogy with waves propagating through dielectric materials, we interpret this phenomenon as the presence of a diagravitational medium for GWs.
Through this medium the wave has a phase velocity,  which is the speed of the phase of the monochromatic GW, and a group velocity given by
\begin{equation}\label{vp}
V_{\rm p}=\frac{1}{n}=c_T+\frac{m_{\rm g}^2}{2\,\omega^2}+\frac{A}{2}\omega^{\alpha-2},
\end{equation} 
and
\begin{equation}\label{vg}
V_{\rm g}=\frac{{\rm d}\omega}{{\rm d}k}=c_{\rm T}-\frac{m_{\rm g}^2}{2\,\omega^2}+\frac{(\alpha-1)A}{2}\omega^{\alpha-2},
\end{equation}
respectively.
It is easy to understand that the following effects on the propagation of GWs can be obtained:
\paragraph{Subluminal speed} For $c_T\simeq {\rm constant}$, this is the only term from those included in equation (\ref{neff}) that does not induce any spread of the wave packet. Therefore, it corresponds to the simple case of a wave moving through a non-dispersive medium, that is $V_{\rm p}=V_{\rm g}$. This modification on the propagation of GWs can be found, for example, when considering scalar-tensor theories with a kinetic coupling to gravity \cite{Kobayashi:2011nu,Bettoni:2016mij}.
\paragraph{Mass} This is the simplest case of a dispersive medium, as it introduces frequency-dependence. Therefore, each component of the wavepacket having a different frequency will propagate at a different speed $V_{\rm p}$, whereas the graviton propagates at $V_{{\rm g},0}\equiv V_{\rm g}(\omega_0)$ if the wavepacket is sharply peaked around $\omega_0$. 
Different theories can equipe the graviton with a non-vanishing mass, although those avoiding the introduction of ghosts are of particular interest \cite{deRham:2014zqa}.
\paragraph{Attenuation} The term $\nu$ implies that the refractive index has a non-vanishing imaginary part.
Light is absorbed by the medium for $\nu>0$ and amplified for $\nu<0$, and the wave packet is spread. In the context of GWs in ATGs, such amplification was first discussed in scalar-tensor theories of gravity \cite{Barrow:1993ad}.  This attenuation seems typically linked to the existence of an effective non-constant gravitational coupling  \cite{Saltas:2014dha} and, just considering a formal and dimensional argument, one could think that it has to be always based on the variation of some fundamental quantity of the ATG.
\paragraph{Lorentz violation} Modified dispersion relations of the form $\omega^2=k^2+A\,k^\alpha$, where $\alpha\neq0,\,2$ and $A\neq0$, are commonly found encapsulating the quantum gravitational phenomenology of different theories \cite{Mirshekari:2011yq,Yunes:2016jcc}. 
The term $A\,k^\alpha$ produces an effect qualitatively similar to that discussed in the previous paragraph, although for the most common cases with $\alpha>1$ the correction on the phase and group velocities have the same sign. For that case, one needs $A<0$ to avoid problematic superluminalities.
Discussions about how to constrain the group velocity (and, therefore, $A$) of these theories using GWs can be found, for example, in references \cite{Yunes:2016jcc,Sotiriou:2017obf}.
\paragraph{Birefrigence} One could further generalize this scenario considering that each polarization ($h_{\times,+}$) propagates according to a different refractive index ($n_{\times,+}$) and with different velocities. This effect appears, for example, in ATGs that induce parity violations \cite{Alexander:2004wk,Yunes:2010yf}.
\paragraph{GW oscillation} In ATGs with more than one dynamical metric, the tensor perturbations are coupled. Linear combinations of those perturbations propagate as if they were in media described by refractive indexes of the form (\ref{neff}), which are not corresponding to the gravitational medium where we measure.
This phenomenon, which is similar in nature to neutrino oscillation, has been studied in detail in bigravity \cite{Max:2017flc} and it can also appear in other theories \cite{Caldwell:2016sut,BeltranJimenez:2018ymu}.

\vspace{0.2cm}
We expect that any ATG will predict GWs that propagate in Minkowski as if they were moving through a medium with a refractive index with a form qualitatively similar to that discussed here. Thus, the presented arguments would apply to any possible ATG, although we can think that the particular dependence of the attenuation on the frequency term may change for unexplored theories.

On the other hand, when considering GWs in more general backgrounds, one should first note that those backgrounds must have two separate scales to allow the definition of GWs. Moreover, in order to consider a refractive index for the diagravitational medium, the symmetry of the spacetime should suggest a natural choice of a time coordinate and, therefore, of a frequency and a wave number vector. 
For example, it is well known that both conditions are satisfied in cosmological homogeneous and isotropic scenarios \cite{Caprini:2018mtu}. In these scenarios, the propagation of GWs is also given by equation (\ref{eqS}) but changing $\nu$ by $2\mathcal{H}+\nu$, with $\mathcal{H}$ being the Hubble rate in conformal time  \cite{Saltas:2014dha}. Hence the diagravitational medium is also isotropic and, therefore, characterized by a scalar effective refractive index of the form given by equation (\ref{nt}) (or equation (\ref{neff}) in the limit $n\simeq1$) with $\nu\rightarrow2\mathcal{H}+\nu$.

\section{Dispersion in anisotropic media}
It is known that one can define an effective refractive index $3\times3$-tensor for electromagnetic waves propagating in curved geometries  \cite{Boonserm:2004wp}. In order to understand the most general form that such a tensor could have when one includes also additional gravitational degrees of freedom, we can look at propagation of light through different kind of materials as a source of inspiration. Propagation of electromagnetic waves through anisotropic dielectric media characterized by two generic permittivity and permeability $3\times3$-matrices, denoted by $\varepsilon^{ij}$ and $\mu^{ij}$, has been studied in reference~\cite{Itin:2009fj}. There the author obtained a compact dispersion relation for electromagnetic waves in an anisotropic medium investigating the existence of solutions for the wave propagation system \cite{Itin:2009fj}
\begin{equation}\label{EM}
\chi^{\alpha\beta\gamma\delta}\partial_\beta\partial_\delta A_\gamma=0,
\end{equation}
where $A_\gamma$ is the vector potential of the electromagnetic field strength and $\chi^{\alpha\beta\gamma\delta}$ is the electromagnetic constitutive tensor that is antisymmetric under permutations of indexes $\alpha$ and $\beta$, and $\gamma$ and $\delta$; and it is symmetric under permutation of $\beta$ and $\delta$. The non-vanishing components of $\chi^{\alpha\beta\gamma\delta}$ can be expressed in terms of $\varepsilon^{ij}$ and $\mu^{ij}$. That dispersion relation is 
\begin{equation}
\omega^4-2\,k_i\,\psi^{i}_j \,k^j\,\omega^2+k_ik_j\,\gamma^{ij}_{mn}\,k^mk^n=0,
\end{equation}
where
\begin{equation}
\psi^{ij}=\frac{1}{2}\epsilon^{imn}\epsilon^{jpq}(\varepsilon^{-1})_{nq}(\mu^{-1})_{mp},
\end{equation}
and
\begin{equation}\label{Itin}
\gamma^{ijmn}=\frac{\varepsilon^{ij}}{{\rm det}\,\varepsilon}\frac{\mu^{mn}}{{\rm det}\,\mu}.
\end{equation}
As it has been proven in reference~\cite{Itin:2009fj}, one recovers the ordinary dispersion relation $k^2= \varepsilon\,\mu\, \omega^2$ from (\ref{Itin}) in the isotropic case $\varepsilon_{ij}=\varepsilon\,\delta_{ij}$ and $\mu_{ij}=\mu\,\delta_{ij}$.
Note that the refractive index is defined in isotropic spaces as $n^2=\varepsilon\,\mu$. Nevertheless, the naive definition of the refractive index tensor based on using square root matrices, that is $n^{ij}n_{jm}=\varepsilon^{ij}\mu_{jm}$, would not allow us to simplify the dispersion relation (\ref{Itin}) as depending only on this $n_{ij}$ in general. 

Now let us go back to the study of GWs predicted by ATGs and propagating in curved background geometries.
We can consider the case in which the propagation equation for GWs can be written as
\begin{equation}\label{eqG}
\Theta^{\alpha\beta\gamma\delta\mu\nu}\partial_\gamma\partial_\delta h_{\mu\nu}=0,
\end{equation}
where we have introduced the {\it gravitational constitutive tensor} $\Theta^{\alpha\beta\gamma\delta\mu\nu}$, which characterizes the gravitational theory in a particular background environment. It is symmetric under permutation of indexes $\alpha$ and $\beta$; $\gamma$ and $\delta$; and $\mu$ and $\nu$, and the pair of indexes $\alpha,\,\beta$ and $\mu,\,\nu$ (see discussion at the end of this section).
We are interested in physically non-trivial solutions of this linear system under analogous conditions of the standard geometric optics approximation applied to the gravitational radiation (\emph{gravito-optics}).
As in the more simple case discussed in the previous section, we will assume a low variation of the entries of the constitutive tensor relative to the change of the metric perturbations. Equation (\ref{eqG}) strongly resembles the source-free wave equation (\ref{EM}), so one could expect that a solution of this equation exists for
\begin{equation}
\label{eq:dispersion1}
\omega^4-2 \, k_{i}\, {{\Psi}^{i}}_{j}\,k^{j} \omega^2
+ k_{i}k_{j}\, {\Gamma^{ij}}_{mn} \,k^{m}k^{n} =0,
\end{equation}
where $\Psi^i{}_j$ and ${\Gamma^{ij}}_{mn}$ are determined by the constitutive tensor and are in general two independent tensors characterizing the medium. 
However, in the case that $\Psi^i{}_j$ and ${\Gamma^{ij}}_{mn}$ are not independent, a unique refractive index tensor can encapsulate the physics of our diagravitational medium. 
For example, when ${\Gamma^{ij}}_{mn}=\Psi^i{}_m\Psi^j{}_n$, the dispersion relation (\ref{eq:dispersion1}) reduces to the quadratic equation
\begin{equation}
k_i\,(n^{-2})^i{}_j\, k^j=\omega^2,
\end{equation}
where we have defined $(n^{-2})^i{}_j \equiv \Psi^i{}_j$.
In the isotropic case, where $n^i{}_j=n\,\delta^i{}_j$, we recover equation (\ref{eqn}).

For illustration, we can find a particular example of non-trivial constitutive tensors focusing our attention on local propagation of GWs in scalar-tensor theories of gravity with a quartic shift-symmetric Horndeski. 
That is
\begin{equation}
L= G( X)\,R+G'(X)\left[\left(\Box\phi\right)^2-\nabla_\mu\nabla_\nu\phi \nabla^\mu\nabla^\nu\phi\right],
\end{equation}
where $G(X)$ is a function of the kinetic term of the scalar field $X=-\partial_\mu\phi\partial^\mu\phi/2$.
The quadratic Lagrangian for tensor perturbations presented in reference \cite{Bettoni:2016mij} can be written as
\begin{equation}
L=\frac{1}{2}h_{\alpha\beta}\,\mathcal{G}^{\gamma\sigma}\partial_\gamma\partial_\sigma h^{\alpha\beta}
+h_{\alpha\beta}\,\mathcal{F}^{\alpha\sigma\gamma\delta}\partial_\gamma\partial_\delta h_\sigma{}^\beta,
\end{equation}
with
\begin{equation}
\mathcal{G}^{\gamma\sigma}=G(X) \,g^{\gamma\sigma}+G'(X)\nabla^\gamma\phi\nabla^\sigma\phi,
\end{equation}
as defined in reference \cite{Bettoni:2016mij}, and 
\begin{equation}
\mathcal{F}^{\alpha\sigma\gamma\delta}=G'(X)\nabla^\alpha\phi\nabla^\sigma\phi \, g^{\gamma\delta}.
\end{equation}
Considering that the entries of the constitutive tensor are slowly varying, 
the equations of motion are given by equation (\ref{eqG}) with
\begin{eqnarray}
\Theta^{\alpha\beta\gamma\delta\mu\nu}=g^{\alpha\mu}g^{\beta\nu}\mathcal{G}^{\gamma\delta}
+2\,g^{\beta\nu}\mathcal{F^{\alpha\mu\gamma\delta}}.
\end{eqnarray}

The picture we have just discussed may be seen being very general. Nevertheless, it should be noted that equation (\ref{eqG}) and, therefore, equation (\ref{eq:dispersion1}) cannot describe all the effects discussed in the previous section.
A possible extension of equation (\ref{eqG}) is
\begin{equation}\label{eqG2}
\left[\Theta^{\alpha\beta\gamma\delta\mu\nu}\partial_\gamma\partial_\delta+\Phi^{\alpha\beta\gamma\mu\nu}\partial_\gamma+\Sigma^{\alpha\beta\mu\nu}\right] \,h_{\mu\nu}=0,
\end{equation}
which includes two additional constitutive tensors, these are the attenuation tensor and the mass tensor. 
The attenuation term appears easily when one changes partial derivatives by covariant derivatives in equation (\ref{eqG}).
It is worth noting that it corresponds with a real attenuation for temporal derivatives. However, for spatial derivatives one would have a contribution of different nature; thus, this term can break the commonly assumed reflexion invariance of the dispersion relation. The origin of the mass term can be traced back to a quadratic term for $h$ in the perturbed Lagrangian. That is
\begin{eqnarray}\label{Lgral}
L&=&-\frac{1}{2}\partial_\gamma h_{\alpha\beta} \,\Theta^{\alpha\beta\gamma\delta\mu\nu}\,\partial_\delta h_{\mu\nu}
+\frac{1}{2}h_{\alpha\beta}\,\Phi^{\alpha\beta\gamma\mu\nu}\partial_\gamma h_{\mu\nu}\nonumber\\
&+&\frac{1}{2}h_{\alpha\beta}\,\Sigma^{\alpha\beta\mu\nu}h_{\mu\nu},
\end{eqnarray}
where we have already discussed the symmetric properties of $\Theta^{\alpha\beta\gamma\delta\mu\nu}$; $\Phi^{\alpha\beta\gamma\mu\nu}$ is antisymmetric under the change of the pair $\alpha,\,\beta$ with $\mu,\,\nu$; and, $\Sigma^{\alpha\beta\mu\nu}$ is symmetric under that change of indexes. We have assumed, on one hand, that all the entries of the tensors are slowly varying functions in the interval of interest and, on the other hand, that Lagrangian (\ref{Lgral}) can be obtained from a covariant Lagrangian of an ATG up to second order perturbations.

\section{Discussion}

The development of the field of gravito-optics may allow us to unveil unexplored characteristics of ATGs. In that context, as we have shown, we can learn from our experience dealing with the propagation of electromagnetic waves through dielectric materials to understand better the behaviour of GWs propagating in curved background geometries. Moreover, the developments on the understanding of new materials could even suggest us the formulation of novel theories.

We have to stress that throughout this work we have assumed that the relevant parameters appearing in the propagation equation, or the entries of the constitutive tensors, can be treated as constants. The constitutive equation describes propagation through the diagravitational medium defined with respect to the vacuum case in GR; thus, the constitutive tensors have an emergent nature.
Going beyond the intervals where this approximation can be assumed will complicate the treatment. However, one could think in some situation such that the different spacetime intervals where this approximation is appropriate could be described as disjoint patches, forming a patchwork of diagravitational regions characterized by different effective refractive indexes.  On the other hand, it is worth mentioning that a medium of collisionless particles as cold dark matter can also affect the propagation of GWs and, therefore, contribute to the constitutive tensors \cite{Flauger:2017ged}.

Finally, it is interesting to emphasize that an arbitrary ATG can support not only tensorial degrees of freedom, but also scalar and vector ones.
The vector and tensorial modes will have different polarizations and each one of these polarizations for each mode can have associated its corresponding mass, attenuation, and Lorentz violating terms and, therefore, its respective anisotropic gravitational refractive index tensor. This tensor is not required to be
symmetric nor even invertible. Indeed, for the electromagnetic radiation, negative
permittivities are associated to metamaterials; non-symmetric tensors describe
external fields effects and non-invertible tensors have been used to describe
perfect conductors \cite{Lindell:2005}. Analogous discussions may apply for the phenomenology of GWs predicted by ATGs propagating in general backgrounds.

\section*{Acknowledgments}
We thank J.~M.~Ezquiaga for useful comments. This work is supported by the project FIS2016-78859-P (AEI/FEDER, UE). PMM gratefully acknowledges financial support provided through the 
Research Award L'Or\'eal-UNESCO FWIS (XII Spanish edition).


\begin{thebibliography}{00}

\bibitem{Abbott:2016blz}
  B.~P.~Abbott {\it et al.} [LIGO Scientific and Virgo Collaborations],
  ``Observation of Gravitational Waves from a Binary Black Hole Merger'',
  Phys.\ Rev.\ Lett.\  {\bf 116} (2016) no.6,  061102
  doi:10.1103/PhysRevLett.116.061102
  [arXiv:1602.03837 [gr-qc]].

\bibitem{Will:2014kxa}
  C.~M.~Will,
  ``The Confrontation between General Relativity and Experiment'',
  Living Rev.\ Rel.\  {\bf 17} (2014) 4
 doi:10.12942/lrr-2014-4
  [arXiv:1403.7377 [gr-qc]].

\bibitem{TheLIGOScientific:2017qsa}
  B.~P.~Abbott {\it et al.} [LIGO Scientific and Virgo Collaborations],
  ``GW170817: Observation of Gravitational Waves from a Binary Neutron Star Inspiral'',
  Phys.\ Rev.\ Lett.\  {\bf 119} (2017) no.16,  161101
  doi:10.1103/PhysRevLett.119.161101
  [arXiv:1710.05832 [gr-qc]].

\bibitem{Lombriser:2015sxa}
  L.~Lombriser and A.~Taylor,
  ``Breaking a Dark Degeneracy with Gravitational Waves'',
  JCAP {\bf 1603} (2016) no.03,  031
  doi:10.1088/1475-7516/2016/03/031
  [arXiv:1509.08458 [astro-ph.CO]].
  
\bibitem{Lombriser:2016yzn}
  L.~Lombriser and N.~A.~Lima,
  ``Challenges to Self-Acceleration in Modified Gravity from Gravitational Waves and Large-Scale Structure'',
  Phys.\ Lett.\ B {\bf 765} (2017) 382
  doi:10.1016/j.physletb.2016.12.048
  [arXiv:1602.07670 [astro-ph.CO]].
  
\bibitem{Ezquiaga:2017ekz}
  J.~M.~Ezquiaga and M.~Zumalac\'arregui,
  ``Dark Energy After GW170817: Dead Ends and the Road Ahead'',
  Phys.\ Rev.\ Lett.\  {\bf 119} (2017) no.25,  251304
  doi:10.1103/PhysRevLett.119.251304
  [arXiv:1710.05901 [astro-ph.CO]].
 
\bibitem{Creminelli:2017sry}
  P.~Creminelli and F.~Vernizzi,
  ``Dark Energy after GW170817 and GRB170817A'',
  Phys.\ Rev.\ Lett.\  {\bf 119} (2017) no.25,  251302
  doi:10.1103/PhysRevLett.119.251302
  [arXiv:1710.05877 [astro-ph.CO]].

\bibitem{Sakstein:2017xjx}
  J.~Sakstein and B.~Jain,
  ``Implications of the Neutron Star Merger GW170817 for Cosmological Scalar-Tensor Theories'',
  Phys.\ Rev.\ Lett.\  {\bf 119} (2017) no.25,  251303
  doi:10.1103/PhysRevLett.119.251303
  [arXiv:1710.05893 [astro-ph.CO]].

\bibitem{Baker:2017hug}
  T.~Baker, E.~Bellini, P.~G.~Ferreira, M.~Lagos, J.~Noller and I.~Sawicki,
  ``Strong constraints on cosmological gravity from GW170817 and GRB 170817A'',
  Phys.\ Rev.\ Lett.\  {\bf 119} (2017) no.25,  251301
  doi:10.1103/PhysRevLett.119.251301
  [arXiv:1710.06394 [astro-ph.CO]].
  
\bibitem{Akrami:2018yjz}
  Y.~Akrami, P.~Brax, A.~C.~Davis and V.~Vardanyan,
  ``Neutron star merger GW170817 strongly constrains doubly-coupled bigravity'',
  arXiv:1803.09726 [astro-ph.CO].
  
\bibitem{deRham:2014zqa}
  C.~de Rham,
  ``Massive Gravity'',
  Living Rev.\ Rel.\  {\bf 17} (2014) 7
 doi:10.12942/lrr-2014-7
  [arXiv:1401.4173 [hep-th]].

\bibitem{Saltas:2014dha}
  I.~D.~Saltas, I.~Sawicki, L.~Amendola and M.~Kunz,
  ``Anisotropic Stress as a Signature of Nonstandard Propagation of Gravitational Waves'',
  Phys.\ Rev.\ Lett.\  {\bf 113} (2014) no.19,  191101
 doi:10.1103/PhysRevLett.113.191101
  [arXiv:1406.7139 [astro-ph.CO]].
 
\bibitem{Caprini:2018mtu}
  C.~Caprini and D.~G.~Figueroa,
  ``Cosmological Backgrounds of Gravitational Waves'',
  arXiv:1801.04268 [astro-ph.CO].
 
\bibitem{Boonserm:2004wp}
  P.~Boonserm, C.~Cattoen, T.~Faber, M.~Visser and S.~Weinfurtner,
  ``Effective refractive index tensor for weak field gravity'',
  Class.\ Quant.\ Grav.\  {\bf 22} (2005) 1905
 doi:10.1088/0264-9381/22/11/001
  [gr-qc/0411034].

\bibitem{Szekeres}
P.~Szekeres, ``Linearized gravitation theory in macroscopic media",
Annals Phys. {\bf 64}, 599 (1971).

\bibitem{Peters}
P.~C.~Peters, ``Index of refraction for scalar, electromagnetic and gravitational waves in weak gravitational field", Phys.~Rev.~D {\bf 9} (1974) 2207.

\bibitem{Giovannini:2015kfa}
  M.~Giovannini,
  ``The refractive index of relic gravitons'',
  Class.\ Quant.\ Grav.\  {\bf 33} (2016) no.12,  125002
 doi:10.1088/0264-9381/33/12/125002
  [arXiv:1507.03456 [astro-ph.CO]].

\bibitem{Itin:2009fj}
  Y.~Itin,
  ``Dispersion relation for anisotropic media'',
  Phys.\ Lett.\ A {\bf 374} (2010) 1113
doi:10.1016/j.physleta.2009.12.071
  [arXiv:0908.0922 [physics.class-ph]].

\bibitem{Giovannini:2018zbf}
  M.~Giovannini,
  ``The propagating speed of relic gravitational waves and their refractive index during inflation'',
  arXiv:1803.05203 [gr-qc].

\bibitem{Giovannini:2018nkt}
  M.~Giovannini,
  ``Blue and violet graviton spectra from a dynamical refractive index'',
  arXiv:1805.08142 [astro-ph.CO].

\bibitem{Mirshekari:2011yq}
  S.~Mirshekari, N.~Yunes and C.~M.~Will,
  ``Constraining Generic Lorentz Violation and the Speed of the Graviton with Gravitational Waves'',
  Phys.\ Rev.\ D {\bf 85} (2012) 024041
  doi:10.1103/PhysRevD.85.024041
  [arXiv:1110.2720 [gr-qc]].

\bibitem{Kobayashi:2011nu}
  T.~Kobayashi, M.~Yamaguchi and J.~Yokoyama,
  ``Generalized G-inflation: Inflation with the most general second-order field equations'',
  Prog.\ Theor.\ Phys.\  {\bf 126} (2011) 511
 doi:10.1143/PTP.126.511
  [arXiv:1105.5723 [hep-th]].

\bibitem{Bettoni:2016mij}
  D.~Bettoni, J.~M.~Ezquiaga, K.~Hinterbichler and M.~Zumalacárregui,
  ``Speed of Gravitational Waves and the Fate of Scalar-Tensor Gravity'',
  Phys.\ Rev.\ D {\bf 95} (2017) no.8,  084029
  doi:10.1103/PhysRevD.95.084029
  [arXiv:1608.01982 [gr-qc]].
 
\bibitem{Barrow:1993ad}
  J.~D.~Barrow, J.~P.~Mimoso and M.~R.~de Garcia Maia,
  ``Amplification of gravitational waves in scalar - tensor theories of gravity'',
  Phys.\ Rev.\ D {\bf 48} (1993) 3630
   Erratum: [Phys.\ Rev.\ D {\bf 51} (1995) 5967].
  doi:10.1103/PhysRevD.48.3630, 10.1103/PhysRevD.51.5967
 
 
\bibitem{Yunes:2016jcc}
  N.~Yunes, K.~Yagi and F.~Pretorius,
  ``Theoretical Physics Implications of the Binary Black-Hole Mergers GW150914 and GW151226'',
  Phys.\ Rev.\ D {\bf 94} (2016) no.8,  084002
 doi:10.1103/PhysRevD.94.084002
  [arXiv:1603.08955 [gr-qc]].

\bibitem{Sotiriou:2017obf}
  T.~P.~Sotiriou,
  ``Detecting Lorentz Violations with Gravitational Waves from Black Hole Binaries'',
  Phys.\ Rev.\ Lett.\  {\bf 120} (2018) no.4,  041104
  doi:10.1103/PhysRevLett.120.041104
  [arXiv:1709.00940 [gr-qc]].
 
\bibitem{Alexander:2004wk}
  S.~Alexander and J.~Martin,
  ``Birefringent gravitational waves and the consistency check of inflation'',
  Phys.\ Rev.\ D {\bf 71} (2005) 063526
  doi:10.1103/PhysRevD.71.063526
  [hep-th/0410230].

\bibitem{Yunes:2010yf}
  N.~Yunes, R.~O'Shaughnessy, B.~J.~Owen and S.~Alexander,
  ``Testing gravitational parity violation with coincident gravitational waves and short gamma-ray bursts'',
  Phys.\ Rev.\ D {\bf 82} (2010) 064017
  doi:10.1103/PhysRevD.82.064017
  [arXiv:1005.3310 [gr-qc]].

\bibitem{Max:2017flc}
  K.~Max, M.~Platscher and J.~Smirnov,
  ``Gravitational Wave Oscillations in Bigravity'',
  Phys.\ Rev.\ Lett.\  {\bf 119} (2017) no.11,  111101
doi:10.1103/PhysRevLett.119.111101
  [arXiv:1703.07785 [gr-qc]].

\bibitem{Caldwell:2016sut}
  R.~R.~Caldwell, C.~Devulder and N.~A.~Maksimova,
  ``Gravitational wave-Gauge field oscillations'',
  Phys.\ Rev.\ D {\bf 94} (2016) no.6,  063005
  doi:10.1103/PhysRevD.94.063005
  [arXiv:1604.08939 [gr-qc]].

\bibitem{BeltranJimenez:2018ymu}
  J.~Beltran Jimenez and L.~Heisenberg,
  ``Non-trivial gravitational waves and structure formation phenomenology from dark energy'',
  arXiv:1806.01753 [gr-qc].

\bibitem{Flauger:2017ged}
  R.~Flauger and S.~Weinberg,
  ``Gravitational Waves in Cold Dark Matter'',
  Phys.\ Rev.\ D {\bf 97} (2018) no.12,  123506
  doi:10.1103/PhysRevD.97.123506
  [arXiv:1801.00386 [astro-ph.CO]].
  
\bibitem{Lindell:2005}
  I.~V. Lindell, A.~Sihvola,
  ``Perfect electromagnetic conductor'',
  Journal of Electromagnetic Waves and Applications {\bf 19},7 (2005) 861-869,
  [arXiv:physics/0503232 [physics.class-ph]].

\end{thebibliography}
\end{document}